\documentclass[twocolumn,aps,prb,citeautoscript,floats,floatfix,superscriptaddress]{revtex4-1}
\usepackage{graphicx}
\usepackage{dcolumn}
\usepackage{amsmath}
\usepackage{amsfonts}
\usepackage{amssymb}
\usepackage{epsfig,float,afterpage,wrapfig,psfrag}
\usepackage{natbib}

\newcommand{\beq}{\begin{equation}}
\newcommand{\eeq}{\end{equation}}
\newcommand{\bes}{\begin{subequations}}
\newcommand{\ees}{\end{subequations}}
\newcommand{\bea}{\begin{eqnarray}}
\newcommand{\eea}{\end{eqnarray}}
\newcommand{\ba}{\begin{array}}
\newcommand{\ea}{\end{array}}
\newcommand{\beqn}{\begin{eqnarray*}}
\newcommand{\eeqn}{\end{eqnarray*}}

\newcommand{\f}[2]{\frac{#1}{#2}}
\newcommand{\g}{\gamma}

\newcommand{\n}{\eta}
\newcommand{\tc}{\tilde{c}}
\newcommand{\tn}{\tilde{\eta}}
\newcommand{\Se}{\Sigma}
\newcommand{\om}{\omega}
\newcommand{\G}{\Gamma}
\newcommand{\la}{\langle}
\newcommand{\ra}{\rangle}

\def\nn{\nonumber}

\newlength{\sizeonefig}
\newlength{\sizetwofig}
\allowdisplaybreaks

\begin{document}

\title{Nature of the zero-bias conductance peak associated with Majorana bound states in topological phases of semiconductor-superconductor hybrid structures}

\author{Dibyendu Roy} 
\affiliation{Theoretical Divisions and Center for Nonlinear Studies, Los Alamos National Laboratory, Los Alamos, NM 87545, USA}
\affiliation{Department of Physics, University of Cincinnati, Cincinnati, Ohio 45221-0011, USA} 

\author{C.~J.~Bolech}
\author{Nayana Shah} 
\affiliation{Department of Physics, University of Cincinnati, Cincinnati, Ohio 45221-0011, USA}

\begin{abstract}
Rashba spin-orbit coupled semiconductor-superconductor hybrid structures in the presence of Zeeman splitting have emerged as the first experimentally realizable topological superconductor supporting zero-energy Majorana bound states. 
However, recent experimental studies in these hybrid structures are not in complete agreement with the theoretical predictions, for example, the observed height of the zero-bias conductance peak (ZBCP) associated with the Majorana bound states is less than $10\%$ of the predicted quantized value $2e^2/h$. 
We try to understand the sources of various discrepancies between the recent experiments and the earlier theories by starting from a microscopic theory and studying non-equilibrium transport in these systems at arbitrary temperatures and applied bias voltages. 
Our approach involves quantum Langevin equations and non-equilibrium Green's functions. Here we are able to model the tunnel coupling between the one-dimensional semiconductor-superconductor hybrid structure and the metallic leads realistically; study the role of tunnel coupling on the height of the ZBCP and the subgap conductance; predict the nature of the splitting of the ZBCP with an increasing magnetic field beyond the critical field; show the behavior of the ZBCP with an increasing gate-controlled onsite potential; and study the evolution of the full differential conductance across the topological quantum phase transition. 
When the applied magnetic field is quite large compared to the Rashba splitting and the bulk energy gap is much reduced, we find the ZBCP even for an onsite potential much larger than the applied magnetic field. The height of the corresponding ZBCP depends on the tunnel coupling even at zero temperature and can be much smaller than $2e^2/h$.
\end{abstract}

\vspace{0.5cm}
\date{\today}

\maketitle

\section{Introduction}

Low-dimensional conventional-superconductor setups have attracted growing attention recently, particularly in the context of the study of fluctuations \cite {nanowires1} and quantum phase transitions \cite{nanowires2}.
On the other hand, Kitaev \cite{Kitaev00} was the first to propose the idea of realizing Majorana fermions as localized states at the ends of an ideal 1D (spinless) $p$-wave superconducting wire motivated by the search for robust quantum information storage. 
This, eventually, generated a growing interest to produce emergent Majorana fermions in the solid-state laboratory setting \cite{Wilczek, Nayak, Beenakker,Alicea12, Stanescu13}. 
A particularly promising proposal to engineer effective $p$-wave superconductivity is by using a Rashba spin-orbit coupled semiconductor in proximity with a conventional $s$-wave (spin-singlet) superconductor and in the presence of Zeeman splitting due to an applied magnetic field \cite{Sau10, Alicea10, Lutchyn10, Oreg10}. 
The Rashba coupling creates two helical bands which wind counterclockwise and clockwise. Introduction of the proximity induced superconductivity gives rise $p$$\pm$i$p$ pairing for the two helical bands. 
One then needs to apply a magnetic field to break the time-reversal doubling of the fermionic states on the spin-orbit coupled semiconductor. 
This thus leads to a topologically nontrivial single species $p$+i$p$ superconductor as originally conceived by Kitaev \cite{Kitaev00}. 
In the presence of a magnetic field, there is an effective $s$-wave pairing between the two helical bands apart from the $p$$\pm$i$p$ pairing for each band. 
The strength of the $p$-wave component of pairing depends on the amount of Rashba splitting. 
Parameters are to be chosen so as to keep a finite $p$-wave component of the induced pairing gap such that the Majorana end states in a wire geometry are protected from the bulk excitations.  

The recent experimental efforts to detect Majorana quasiparticles in one-dimensional semiconductor-superconductor hybrid structures in the presence of strong spin-orbit coupling and a Zeeman field \cite{Mourik12,Deng12,Das12,Churchill13} display several discrepancies with the existing theoretical predictions \cite{Lutchyn10,Oreg10}. 
These experimental studies have mostly concentrated in measuring a zero-bias conductance peak in the tunneling differential conductance of hybrid structures at low temperatures by tuning the applied magnetic field along the one-dimensional (1D) structure. 
It has been theoretically predicted that the existence of zero-energy Majorana bound states (MBSs) in these systems manifests through a quantized zero-bias conductance peak (ZBCP) of height $2e^2/h$ at zero-temperature and above a critical magnetic field ($B_c$) which is determined by the gating of the spin-orbit coupled semiconductor and the magnitude of the induced superconducting gap.  
On the other hand, the height of the observed ZBCP in the experiments is much lower (less than 10$\%$) than the theoretical predictions. 
It has been also seen that the measured ZBCP splits at higher magnetic fields beyond $B_c$. There are also additional discrepancies between theory and experiments which are not yet completely resolved. 
For example, it was theoretically predicted that the superconducting gap would close across a magnetic-field or onsite-potential driven topological quantum phase transition between a non-topological ($s$-wave) superconducting phase and a topological ($p$-wave) superconducting phase with zero-energy Majorana modes \cite{Lutchyn10,Sau10,Alicea10,Stanescu12}. However, some experiments \cite{Das12} have claimed to observe such gap closing and some have not seen a gap closing \cite{Mourik12}. The expected disappearance of the ZBCP as the Zeeman field is rotated from the wire axis in the plane formed by the wire axis and the direction of the effective spin-orbit field has not been observed in the latest experiment \cite{Churchill13}, but it was seen in the previous experiments \cite{Mourik12,Das12}.  
The emergence of real MBSs in these systems can also be probed by looking for a fractional Josephson effect in some type of interference experiments \cite{Kitaev00,Kwon04,Alicea11,Dominguez12}, which manifests though a $4\pi$ periodicity in an ac Josephson measurement. 
These type of doubled-Shapiro-step measurements \cite{Rokhinson12} are complex and harder to interpret theoretically with a realistic microscopic modeling.

The existing theoretical studies to calculate current-voltage characteristics of various topological-superconductor configurations use tools such as formalisms based on the Landauer-B{\"u}ttiker scattering theory \cite{Anantram96, Akhmerov11, Lin12, Beenakker}, or the Keldysh nonequilibrium Green's function (NEGF) formalism \cite{Bolech07, Flensberg10} (as well as other variant approaches, see for instance Ref.~\onlinecite{Prada12}). 
With the latter, models have been more idealized, while with the former, models do not take into account the bath or the coupling with it explicitly but in an effective manner. 
While a bath can induce decoherence in the semiconductor-superconductor hybrid structures, (and thus limit the performance as a qubit of a pair of MBSs), its coupling to the hybrid structures also controls the height of the ZBCP at finite temperature. 
Therefore, an explicit modeling of the baths and their couplings is an important issue to understand the recent experimental data. 
Recently, we have extended the quantum Langevin equations and Green's functions (LEGF) formalism to study nonequilibrium transport in $p$-wave superconductors \cite{Roy12}. 
One advantage of the LEGF method is that it starts with an explicit Hamiltonian for the bath and the coupling along with the system Hamiltonian, thus the role of the bath and the coupling is quite clearly elucidated. 
In this paper we provide a detail derivation of the LEGF method and further extend it for the case of spin-orbit coupled semiconductor-superconductor heterostructures in the presence of an external magnetic field. 
We demonstrate a nonmonotonic dependence of the height of the ZBCP with the applied magnetic field and the onsite potential. 
We also show the oscillation of the splitting width in the ZBCP with an increasing magnetic field beyond $B_c$. 
We further show how the increase of the tunnel coupling can modify the height of the ZBCP at finite temperatures and control the appearance of sub-gap conductance. 
In short, we provide a complete theory of linear and nonlinear transport in the topological superconductors at all temperatures and elucidate many interesting results for the dependence of the ZBCP on various experimentally tunable parameters. 
The rest of this paper is organized as follows: in Sec.~\ref{sec:model} we introduce the model of the wire, in Sec.~\ref{sec:LEGF} we explain the treatment using quantum Langevin equations and Green's functions, and in Sec.~\ref{sec:ZBCP} we present the results mentioned above.

\section{Model and Majorana modes}
\label{sec:model}
Let us consider a single-channel semiconductor nanowire with a strong Rashba spin-orbit coupling (for example, as in InSb or InAs) in close proximity to an ordinary ($s$-wave) superconductor (such as, for instance, NbN, NbTiN, or Al) and in the presence of an external magnetic field applied along the axis of the nanowire. It has been theoretically predicted earlier \cite{Sau10, Lutchyn10, Oreg10} that this system undergoes a topological quantum phase transition at a certain critical magnetic field, $B_c=\sqrt{\Delta^2+\mu^2}$ where $\Delta$ is a proximity induced superconducting gap and $\mu$ is an onsite potential for the nanowire. For an applied magnetic field $B>B_c$ the hybrid structure is driven into a chiral $p$-wave topological superconducting phase supporting two real zero-energy MBSs at the two ends of the nanowire. The requirement of finite $p$-wave component of induced pairing gap for the protection of the MBSs along with the relation for the critical field $B_c$ impose a stringent restriction on the onsite potential or carrier density. One needs a low carrier density in order to satisfy those conditions, therefore the engineered structures are susceptible to disorder. The ZBCP in the differential conductance of the tunneling transport calculations is a signature of the emergence of MBSs in the topological superconducting phase. The semiconducting wire in the semiconductor-superconductor heterostructure is modeled by a tight-binding lattice Hamiltonian of electrons with a proximity induced s-wave BCS pairing of amplitude $\Delta$ \cite{Lutchyn10, Oreg10}. The wire has $N$ lattice sites and each of its two end sites is coupled to an infinite metallic bath which is itself modeled by a one-dimensional tight-binding system of free electrons. The effective Hamiltonian of the system consisting of the nanowire \cite{Sticlet12, Potter11}, the baths, and the tunnel couplings reads  
\begin{eqnarray}
\mathcal{H}&=&\mathcal{H}_{W} +\mathcal{H}_B^L+\mathcal{H}_B^R+\mathcal{H}_{WB}^L+\mathcal{H}_{WB}^R,\label{Ham}\\
\mathcal{H}_{W}&=&\sum_{l=1}^{N-1}\big[-\gamma\sum_{\sigma=\uparrow \downarrow}(a^{\dag}_{l,\sigma} a_{l+1,\sigma}+ a^{\dag}_{l+1,\sigma} a_{l,\sigma})\nn\\&+&\alpha(a^{\dag}_{l+1,\uparrow}a_{l,\downarrow}-a^{\dag}_{l+1,\downarrow}a_{l,\uparrow}+a^{\dag}_{l,\downarrow}a_{l+1,\uparrow}-a^{\dag}_{l,\uparrow}a_{l+1,\downarrow})\big]\nn\\&+&2(\mu-\gamma)\sum_{\sigma=\uparrow \downarrow,l=1}^{N} (a^{\dag}_{l,\sigma}a_{l,\sigma}-\f{1}{2})\nn\\&+&\sum_{l=1}^{N}\big[2B(a^{\dag}_{l,\uparrow}a_{l,\downarrow}+a^{\dag}_{l,\downarrow}a_{l,\uparrow})-2\Delta(a^{\dag}_{l,\uparrow}a^{\dag}_{l,\downarrow}+a_{l,\downarrow} a_{l,\uparrow}) \big],\nn \\
\mathcal{H}_B^m &=&\sum_{\sigma=\uparrow \downarrow,n=1}^{\infty}-\gamma_m(a^{m\dag}_{\sigma,n}a^m_{\sigma,n+1}+a^{m\dag}_{\sigma,n+1} a^m_{\sigma,n}),~m=L,R, \nn \\
\mathcal{H}_{WB}^m&=&-\gamma'_m\sum_{\sigma=\uparrow \downarrow} (a^{m\dag}_{\sigma,1} a_{\sigma,l_m} + a^{\dag}_{\sigma,l_m} a^m_{\sigma,1} ),~l_L=1,l_R=N.\nn
\end{eqnarray}
Here $a^{\dag}_{l,\sigma}$ and $a^{m\dag}_{n,\sigma}$ denote respectively an electron creation operator on the semiconductor nanowire and on the $m^{th}$ bath (here $m=L,R$). The Hamiltonian of the nanowire is  denoted by $\mathcal{H}_W$,  that of the $m^{\rm th}$ bath by $\mathcal{H}^m_B$, and the tunnel coupling between the nanowire and the $m^{\rm th}$ bath is $\mathcal{H}^m_{WB}$. The hopping amplitude of an electron on the wire is $\gamma$, $\mu$ is an onsite potential energy of the wire controlled by gating, $\alpha$ is the strength of the Rashba spin-orbit coupling on the semiconductor wire and $B$ is an applied magnetic field along the axis of the wire.  We shall choose below the values of the different parameters for our plots to be consistent with the values reported in recent experiments.

Next we introduce the following  local transformation for the electron operator of the semiconductor nanowire,
\bea
a_{l,\sigma}&=&\f{1}{2}(c_{A,l,\sigma}+i~c_{B,l,\sigma}),~{\rm with}\nn\\
c^{\dag}_{\beta,l,\sigma}&=&c_{\beta,l,\sigma},~\{c_{\beta,l,\sigma},c_{\beta',l',\sigma'}\}=2\delta_{\beta,\beta'}\delta_{l,l'}\delta_{\sigma,\sigma'},
\eea
where $c_{\beta,l,\sigma}$ is a Majorana (or real) fermion operator. The semiconductor-superconductor hybrid structure Hamiltonian in the Majorana-fermion basis is thus rewritten as:
\begin{widetext}
\bea
\mathcal{H}_W &=&\f{i}{2}\Big[-\gamma\sum_{\sigma=\uparrow \downarrow,l=1}^{N-1}(c_{A,l,\sigma}c_{B,l+1,\sigma}+c_{A,l+1,\sigma}c_{B,l,\sigma})+2(\mu-\gamma) \sum_{\sigma=\uparrow \downarrow,l=1}^N c_{A,l,\sigma}c_{B,l,\sigma}+\alpha\sum_{l=1}^{N-1}(c_{A,l,\downarrow} c_{B,l+1,\uparrow}\nn\\&+&c_{A,l+1,\uparrow} c_{B,l,\downarrow}-c_{A,l,\uparrow} c_{B,l+1,\downarrow}-c_{A,l+1,\downarrow} c_{B,l,\uparrow})+\sum_{l=1}^{N}\big[2(B+\Delta)c_{A,l,\uparrow} c_{B,l,\downarrow}+2(B-\Delta)c_{A,l,\downarrow} c_{B,l,\uparrow}\big]\Big].\label{HamM}
\eea
\end{widetext}

\subsection{Low Energy Spectrum}

When $B>B_c$ we find that the above $4N\times 4N$ matrix of the Hamiltonian in Eq.~(\ref{HamM}) has two degenerate zero-energy eigenstates for large $N$. These zero-energy Majorana modes are separated from all the other energy eigenvalues by an energy gap of order $\Delta$. However, there is a splitting between the two Majorana modes for smaller values of $N$ and the energy of the Majorana modes is not exactly zero. The crossover between these two, longer- \textit{vs} smaller-length, behaviors of the MBSs is determined by the coherence length $\xi_0$ (which is roughly given by $\pi v_F/\Delta$, where $v_F$ is the Fermi velocity) of the spin-orbit coupled semiconductor-superconductor hybrid structure. The coherence length is longer for a smaller proximity-induced superconducting gap or spin-orbit coupling, and it also depends on the applied magnetic field and the onsite potential \cite{DasSarma12}. It is believed that the length of the nanowires in recent experiments was of the same order as the coherence length $\xi_0$ \cite{Mourik12, DasSarma12}. Therefore it is expected that there is an overlap between the two MBSs at the two ends of the nanowire in these experiments. The overlap between the MBSs at the two ends goes to zero for a long nanowire where the length of the wire is much longer than the coherence length $\xi_0$. Whenever there is splitting in the energy of the MBSs of an isolated hybrid structure, it is expected to show up as a splitting in the ZBCP of tunneling measurements as long as the broadening of the conductance peak due to the coupling with the bath and the temperature is smaller than the energy splitting of the two MBSs. We also find that as the applied magnetic field is increased much above $B_c$, the bulk energy gap is substantially reduced. There are still zero-energy real eigenstates at these parameters, however these zero-energy states are very delocalized and extend into the bulk of the wire. When the onsite potential is increased at a higher magnetic field, we find zero-energy states in the isolated Hamiltonian of Eq.~(\ref{HamM}) even when the applied field becomes smaller compared to the critical field at that value of the potential (cf.~\cite{Kells12}). But when the onsite potential is substantially increased the spectrum of the Hamiltonian in Eq.~(\ref{HamM}) becomes again gaped near zero energy, and there is no zero-energy state (same as for the simpler Kitaev model). Below, we consider this regime of parameter sets carefully in our analysis of transport, because we suspect that the recent experiments might be probing it. Although the spectrum and local density of states of this regime had already been analyzed via exact diagonalization of the isolated-wire Hamiltonian, a transport analysis had not been carried out previously.   

\section{Quantum Langevin equations and Green's function formalism}
\label{sec:LEGF}
Here we develop a steady-state non-equilibrium transport theory for the spin-orbit coupled semiconductor-superconductor hybrid structures by employing quantum Langevin equations and Green's functions. We obtain a set of generalized quantum Langevin equations of motion for the nanowire's operators in the Majorana basis following Refs.~\cite{Dhar03, DharSen06, DharRoy06, Roy07}. The description here for the hybrid structures is similar to that for the Kitaev chain in our earlier study \cite{Roy12}. We assume that the metallic baths at the two ends of the nanowire are disconnected from the wire for all times $t \leq t_0$. Each bath is in thermal equilibrium at a specified temperature $T_m$ and a chemical potential $\mu_m$ for both spin components of the bath. We connect the baths to the nanowire at a time $t_0$, and we are interested in the steady-state properties of the nanowire that set in after the end of any transient dynamics. We write the Heisenberg equations of motion for the operators on the nanowire in the Majorana basis at $t > t_0$,  
\begin{widetext}
\begin{eqnarray}
\dot{c}_{A,l,\sigma}&=&\f{2(\mu-\gamma)}{\hbar}c_{B,l,\sigma}+ \f{2(B+\Delta)}{\hbar}\delta_{\sigma,\uparrow}c_{B,l,\downarrow}+\f{2(B-\Delta)}{\hbar}\delta_{\sigma,\downarrow}c_{B,l,\uparrow}-\f{\gamma}{\hbar}c_{B,l+1,\sigma}-\f{\gamma}{\hbar}c_{B,l-1,\sigma}+\f{\alpha}{\hbar}\delta_{\sigma,\downarrow}c_{B,l+1,\uparrow}\nn\\&+&\f{\alpha}{\hbar}\delta_{\sigma,\uparrow}c_{B,l-1,\downarrow}-\f{\alpha}{\hbar}\delta_{\sigma,\downarrow}c_{B,l-1,\uparrow}-\f{\alpha}{\hbar}\delta_{\sigma,\uparrow}c_{B,l+1,\downarrow}+ \f{i\g_L'}{\hbar}\delta_{l,1}(a^L_{1,\sigma}-a^{L\dag}_{1,\sigma})+ \f{i\g_R'}{\hbar}\delta_{l,N}(a^R_{1,\sigma}-a^{R\dag}_{1,\sigma}),  \label{eqmW1}\\
\dot{c}_{B,l,\sigma}&=&-\f{2(\mu-\gamma)}{\hbar}c_{A,l,\sigma}- \f{2(B-\Delta)}{\hbar}\delta_{\sigma,\uparrow}c_{A,l,\downarrow}-\f{2(B+\Delta)}{\hbar}\delta_{\sigma,\downarrow}c_{A,l,\uparrow}+\f{\gamma}{\hbar}c_{A,l+1,\sigma}+\f{\gamma}{\hbar}c_{A,l-1,\sigma}-\f{\alpha}{\hbar}\delta_{\sigma,\downarrow}c_{A,l+1,\uparrow}\nn\\&-&\f{\alpha}{\hbar}\delta_{\sigma,\uparrow}c_{A,l-1,\downarrow}+\f{\alpha}{\hbar}\delta_{\sigma,\downarrow}c_{A,l-1,\uparrow}+\f{\alpha}{\hbar}\delta_{\sigma,\uparrow}c_{A,l+1,\downarrow}+\f{\g_L'}{\hbar}\delta_{l,1}(a^L_{1,\sigma}+a^{L\dag}_{1,\sigma})+\f{\g_R'}{\hbar}\delta_{l,N}(a^R_{1,\sigma}+a^{R\dag}_{1,\sigma}), \label{eqmW2}
\end{eqnarray}
\end{widetext}
for $l = 1,2,3,...,N$ and with the convention that $c_{A,0,\sigma}=c_{A,N+1,\sigma}=c_{B,0,\sigma}=c_{B,N+1,\sigma}=0$. 
The Heisenberg equations of motion for the bath operators (where $\sigma=\uparrow,\downarrow$ and $m=L,R$) are:
\bea
\dot{a}^m_{n,\sigma}&=& \f{i \g_m}{\hbar} (a^m_{n-1,\sigma}+a^m_{n +1,\sigma}),~{\rm for}~n=2,3,...\infty, \label{eqmR1}\\ 
\dot{a}^L_{1,\sigma} &=&\f{i\g_L}{\hbar} a^L_{2,\sigma}+\f{i \g'_L}{\hbar} a_{1,\sigma}, \label{eqmR2}\\ 
\dot{a}^R_{1,\sigma} &=& \f{i \g_R}{\hbar} a^R_{2,\sigma}+ \f{i \g'_R}{\hbar} a_{N,\sigma}. \label{eqmR3}
\eea
The equations of motion of the wire operators, Eqs.~(\ref{eqmW1},\ref{eqmW2}), in the Majorana basis involve the bath variables $a^m_{1,\sigma},a^{m\dag}_{1,\sigma}$ with $m=L,R$, that we can eliminate by replacing with their exact solutions. For that we note that the equations of motion of the each bath, given by Eqs.~(\ref{eqmR1},\ref{eqmR2},\ref{eqmR3}), are a set of linear coupled equations with an inhomogeneous part given by $ i\gamma'_m a_{m,\sigma}/\hbar$. We solve these equations of motion using the single-particle retarded Green's function of the isolated baths, which is given by $g^{m+}_{\sigma}(t)=-i \theta(t) e^{-i H^m_{\sigma} t/\hbar}$ where $H^m_{\sigma}$ is the single-particle Hamiltonian of the spin $\sigma$ component of the $m^{\rm th}$ bath. As here the single particle retarded Green's function of the each bath is the same for both spin components, hereafter we avoid the spin index in $g^{m+}_{\sigma}(t)$, and write it as $g^{m+}(t)$. One finally finds that the solution for the boundary site on the $m^{\rm th}$ bath is given by (for $t > t_0$)
\bea
 a^L_{1,\sigma}(t)&=&i \sum_{n= 1}^{\infty}g^{L+}_{1n}(t-t_0) a^L_{n,\sigma}(t_0)\nn\\ &-&\int_{t_0}^{\infty}dt'~ g^{L+}_{1,1}(t-t')\f{\g'_L}{\hbar} a_{1,\sigma}(t'),\label{LBsol} \\
a^R_{1,\sigma}(t)&=& i \sum_{n = 1}^{\infty}g^{R+}_{1n}(t-t_0)a^R_{n,\sigma}(t_0)\nn\\ &-& \int_{t_0}^{\infty}dt'~ g^{R+}_{1,1}(t-t')\f{\g'_R}{\hbar} a_{N,\sigma}(t').\label{RBsol}
\eea
Plugging these solutions into Eqs.~(\ref{eqmW1},\ref{eqmW2}) for the wire operators in the Majorana basis, we get a set of generalized quantum Langevin equations (see Appendix \ref{GQLE} for the full expressions) where we identify $\eta_{m,\sigma}$ as a noise contribution from the spin-$\sigma$ component of the $m^{\rm th}$ bath while the terms involving $\Sigma_m^\pm(t)$ are the corresponding dissipative terms.
\bea
\n_{m,\sigma}(t) &=&  - \f{i \g_m'}{\hbar} \sum_{n =1}^{\infty}  g^{m+}_{1n}(t-t_0)~a^m_{n,\sigma}(t_0), \\
\Sigma^+_m(t) &=& (\f{\g_m'}{\hbar})^2 g^{m+}_{1,1}(t), ~{\rm and}~~\Sigma^-_m(t)=[\Sigma^+_m(t)]^\dag.
\eea
The noise depends on the bath's initial distribution which we have chosen to correspond to thermal equilibrium. 
The properties of the noise are written most conveniently in the frequency domain. Let us consider the limit $t_0 \to - \infty$, and introduce the Fourier transforms $\tc_{\beta,l,\sigma}(\om)= (1/2\pi) \int_{-\infty}^\infty dt e^{i \om t} c_{\beta,l,\sigma}(t)$, $g^{m+}(\om)=\int_{-\infty}^\infty dt e^{i \om t} g^{m+}(t)$, $\tn_{m,\sigma}(\om)=(1/2\pi)~ \int_{-\infty}^\infty dt
e^{i \om t} \n_{m,\sigma}(t)$ and $\Se^+_m(\om) = (\g'_m/\hbar)^2g^{m+}_{1,1}(\om)$ (here $g^{m+}_{1,1}(\om)$ is the $m^{\rm th}$ bath single-particle Green's function evaluated at the first site). We also have $\tc^{\dag}_{\beta,m,\sigma}(\om)=\tc_{\beta,m,\sigma}(-\om)$. We use the definition $\G_m (\om)=-Im[\Se^+_m]/\pi= (\g'_m/\hbar)^2 \rho_m(\om)$, where $\rho_m (\om)$ is the local density of states of either spin component at the first site ($n=1$) on the $m^{\rm th}$ bath. With these definitions it is easy to show that the noise-noise correlations are given by 
\begin{equation}
\la \tn_{l,\sigma}^\dag (\om) \tn_{m,\sigma'}(\om') \ra = \G_l(\om) f(\omega,\mu_l,
T_l)\delta (\omega -\omega')\delta_{lm}\delta_{\sigma,\sigma'}, \label{nncor}
\end{equation}
where $f(\om,\mu_m,T_m)=1/\{{\rm exp}[(\hbar \om-\mu_m)/k_B T_m]+1\}$ is the Fermi distribution function and $k_B$ is the Boltzmann constant. The eigenvalues and eigenfunctions of the $m^{\rm th}$ bath Hamiltonian are given by $\epsilon^m_q=-2\g_m\cos q$ and $\psi_q(p)=\sqrt{2} \sin (q p)$, where $q$ lies in the range $[0,\pi]$ and $p$ is an integer. By convention, the wave functions are normalized so that $\la \psi_q|\psi_{q'}\ra=\pi \delta(q-q')$. Therefore,
\bea
g^{m+}_{1,1}(t)&=&-i\theta(t)\sum_q \psi_q(1)\psi_q^*(1)e^{-i\epsilon^m_q t/\hbar},\quad\mathrm{and}\nn\\
g^{m+}_{1,1}(\om)&=&\int_{-\infty}^\infty dt~g^{m+}_{1,1}(t)e^{i \om t} \nn\\
&=&{\rm lim}_{\eta\to 0}\sum_q \f{|\psi_q(1)|^2}{\om-\epsilon^m_q/\hbar +i\eta}\nn\\
&=&\sum_q \f{|\psi_q(1)|^2}{\om-\epsilon^m_q/\hbar }-i\pi\sum_q |\psi_q(1)|^2\delta(\om-\epsilon^m_q/\hbar).\nn
\eea  
After converting the $q$ sum to an integral in the range $[0,\pi]$, we find that, within the band-width of the bath ($|\hbar \om |< 2 \gamma_m$), $g_{1,1}^{m+}(\om)$ is given by
\bea
g_{1,1}^{m+}(\om)=\f{\hbar}{\gamma_m}\left[ \f{\hbar \om}{2 \g_m}-i \left(
  1-\frac{\hbar^2 \omega^2}{4 \g_m^2}\right)^{1/2} \right].
\eea
Using these single-particle retarded Green's functions of the baths, we find
\bea
&&\Se^-_m(-\om)- \Se^+_m(\om)=-2\Se^+_m(\om),~\Se^-_m(-\om)+ \Se^+_m(\om)=0,\nn\\&&\f{1}{2i\pi}[\Se^-_m(\om)- \Se^+_m(\om)]=\G_m(\om).
\eea
We solve the generalized quantum Langevin equations in  Eqs.~(\ref{gle1},\ref{gle2}) by Fourier transform and we get the following steady-state solution for the operators on the nanowire.
\bea
\tc_l(\om)&=&\sum_{m=1}^{4N} G^+_{lm}(\om)\tilde{h}_m(\om),~{\rm where} \nn\\
G^+(\om)&=&Z^{-1}(\om),~Z_{lm}(\om)=\Phi_{lm}(\om)+A_{lm}(\om),\label{sol} 
\eea
and we use the following notation for simplicity, $c_{A,l,\uparrow}\equiv c_{4l-3},~c_{B,l,\uparrow}\equiv c_{4l-1},~c_{A,l,\downarrow}\equiv c_{4l-2},~c_{B,l,\downarrow}\equiv c_{4l}$ for $l=1,2,3,...,N$. The expressions for $\Phi_{lm}(\om),~A_{lm}(\om)$ and $\tilde{h}_m(\om)$ are given in Appendix \ref{GreenFn}. Here $G^+(\om)$ is the Green's function of the full system consisting of the nanowire and the baths. We can calculate nonequilibrium steady-state properties of the hybrid structures using the above solutions for the operators. We now proceed in the next section to evaluate the electrical current in the system under an arbitrarily large applied bias voltage and finite temperature.

\section{Current-voltage characteristics and Zero-bias conductance peak}
\label{sec:ZBCP}

We define a local charge density at the boundary site of the nanowire and find an expression for the electrical current through the nanowire using the local density and the continuity equations \cite{Dhar03, Roy07}.  We call $j_{m}(t)$ the inward electrical current flowing from the $m^{\rm th}$ bath into the nanowire. It is given by 
\bea
j_{m}(t)&=&\f{i\gamma_{m}^{\prime}}{\hbar}\langle\sum_{\sigma=\uparrow,\downarrow}(a_{l_m,\sigma}^{\dag}a_{1,\sigma}^{m}-{a_{1,\sigma}^{m}}^{\dag}a_{l_m,\sigma})~\rangle\nn\\
&=&-2~\mathrm{Im}\big[\f{\gamma_{m}^{\prime}}{\hbar}\langle \sum_{\sigma=\uparrow,\downarrow} a_{l_m,\sigma}^{\dag}a_{1,\sigma}^{p}\rangle\big],\label{Ecurr}
\eea
where $\langle...\rangle$ denotes averaging over noise using the result of Eq.~(\ref{nncor}). Notice that, due to the proximity-induced superconductivity in the nanowire, the total electron charge is not conserved. Thus, the electrical current from the left bath into the nanowire is not necessarily equal in magnitude to that from the nanowire into the right bath, for arbitrary chemical potentials in the baths. In order to be specific, from now on we explicitly discuss the electrical current from the left bath. We find from Eq.~\ref{Ecurr}, after using Eq.~\ref{LBsol} with $t_0 \to -\infty$,
\bea
j_{L}(t)&=&2~{\rm Im}\big(\la \sum_{\sigma=\uparrow,\downarrow} 
\big\{a_{1,\sigma}^\dag(t)\big[\eta_{L,\sigma}(t)\nn\\&+&\int_{-\infty}^{\infty}dt'~ \Sigma^+_L(t-t')a_{1,\sigma}(t')\big]\big\}\ra\big).\label{Lcurr}
\eea
Let us derive each part of the above expression separately. After averaging over the noise using Eq.~\ref{nncor}, we find for the first part of Eq.~\ref{Lcurr},
\bea
&&\la\sum_{\sigma=\uparrow,\downarrow}a_{1,\sigma}^\dag(t)\eta_{L,\sigma}(t)\ra=\int_{-\infty}^{\infty}\f{d\omega}{2}\Big\{\big[-G^+_{11}(-\om)-iG^+_{13}(-\om)\nn\\&&+iG^+_{31}(-\om)-G^+_{33}(-\om)\big]\G_L(\om)f(\om,\mu_L,T_L)+[-G^+_{22}(-\om)\nn\\&&-iG^+_{24}(-\om)+iG^+_{42}(-\om)-G^+_{44}(-\om)\big]\G_L(\om)f(\om,\mu_L,T_L)\Big\}.\nn
\eea
After applying the local Majorana basis transformation we find for the second part of Eq.~\ref{Lcurr},
\bea
&&\la~\sum_{\sigma=\uparrow,\downarrow} a_{1,\sigma}^\dag(t)\int_{-\infty}^{\infty}dt' \Sigma^+_L(t-t')a_{1,\sigma}(t')\ra\nn\\
&&=\f{1}{4}\int_{-\infty}^{\infty}dt'\Sigma^+_L(t-t')\la\sum_{\sigma=\uparrow,\downarrow}(c_{A,1,\sigma}(t)c_{A,1,\sigma}(t')+c_{B,1,\sigma}(t)\nn\\&&\times c_{B,1,\sigma}(t')-ic_{B,1,\sigma}(t)c_{A,1,\sigma}(t')+ic_{A,1,\sigma}(t)c_{B,1,\sigma}(t'))\ra,\nn
\eea
in where we can now use the solutions of the operators on the nanowire given by Eq.~\ref{sol}. For example, the first term is given by 
\bea
&&\f{1}{4}\int_{-\infty}^{\infty}dt' \Sigma^+_L(t-t')\la\sum_{\sigma=\uparrow,\downarrow}c_{A,1,\sigma}(t)c_{A,1,\sigma}(t')\ra\nn\\
&&=\f{1}{4}\int_{-\infty}^{\infty}\int_{-\infty}^{\infty}d\om d\om' e^{-i(\om+\om')t}~\Sigma^+_L(\om)\sum_{m=1}^{4N}\sum_{n=1}^{4N}(G^+_{1m}(\om')\nn\\&&\times G^+_{1n}(\om)+G^+_{2m}(\om')G^+_{2n}(\om))\la\tilde{h}_m(\om')\tilde{h}_n(\om)\ra.\label{Lcurr2} 
\eea
The noise average in Eq.~\ref{Lcurr2} is carried out using Eq.~\ref{Noise} along with the noise correlation properties from Eq.~\ref{nncor}. In the steady state, $j_L(t)$ is independent of time. Here we evaluate the current-voltage characteristics of the hybrid nanowire structures in the steady state.  We can calculate the full Green's function in Eq.~\ref{sol} numerically and use them to find the current at zero or finite temperatures using the above results. 

\subsection{Numerical Results}

We now focus on the `symmetric' case for which the two baths are identical ($\gamma_{L}=\gamma_{R}$) and are connected to the nanowire by identical contacts ($\gamma^{\prime}_{L}=\gamma^{\prime}_{R}$ or $\Gamma_{L}=\Gamma_{R}\equiv\Gamma$) while $\mu_{L}=\tilde{\mu}=-\mu_{R}$ and $T_{L}=T=T_{R}$. In this case the steady-state currents $j_{L,R}$ are equal up to a sign and
\begin{equation}
j_{L}(\tilde{\mu},-\tilde{\mu})=\int_{-\infty}^{\infty}\f{d\omega}{2\pi} \mathcal{T}(\om)\big[f(\om,\tilde{\mu},T)-f(\om,-\tilde{\mu},T)\big] \label{lsCurr}
\end{equation}
Here $\mathcal{T}(\omega)$, which contains contributions to transport coming from both electrons and holes, can be simply interpreted in the symmetric case via the zero-temperature differential conductance (with $\tilde{\mu}\equiv eV$, $e=1$)
\begin{equation}
\frac{dI}{dV}=\frac{dj_{L}(\tilde{\mu},-\tilde{\mu})}{d\tilde{\mu}%
}=\frac{1}{2\pi}\Big(\mathcal{T}(\tilde{\mu})+\mathcal{T}(-\tilde{\mu})\Big)
\end{equation}

\begin{figure}
\begin{tabular}{cc}
\epsfig{file=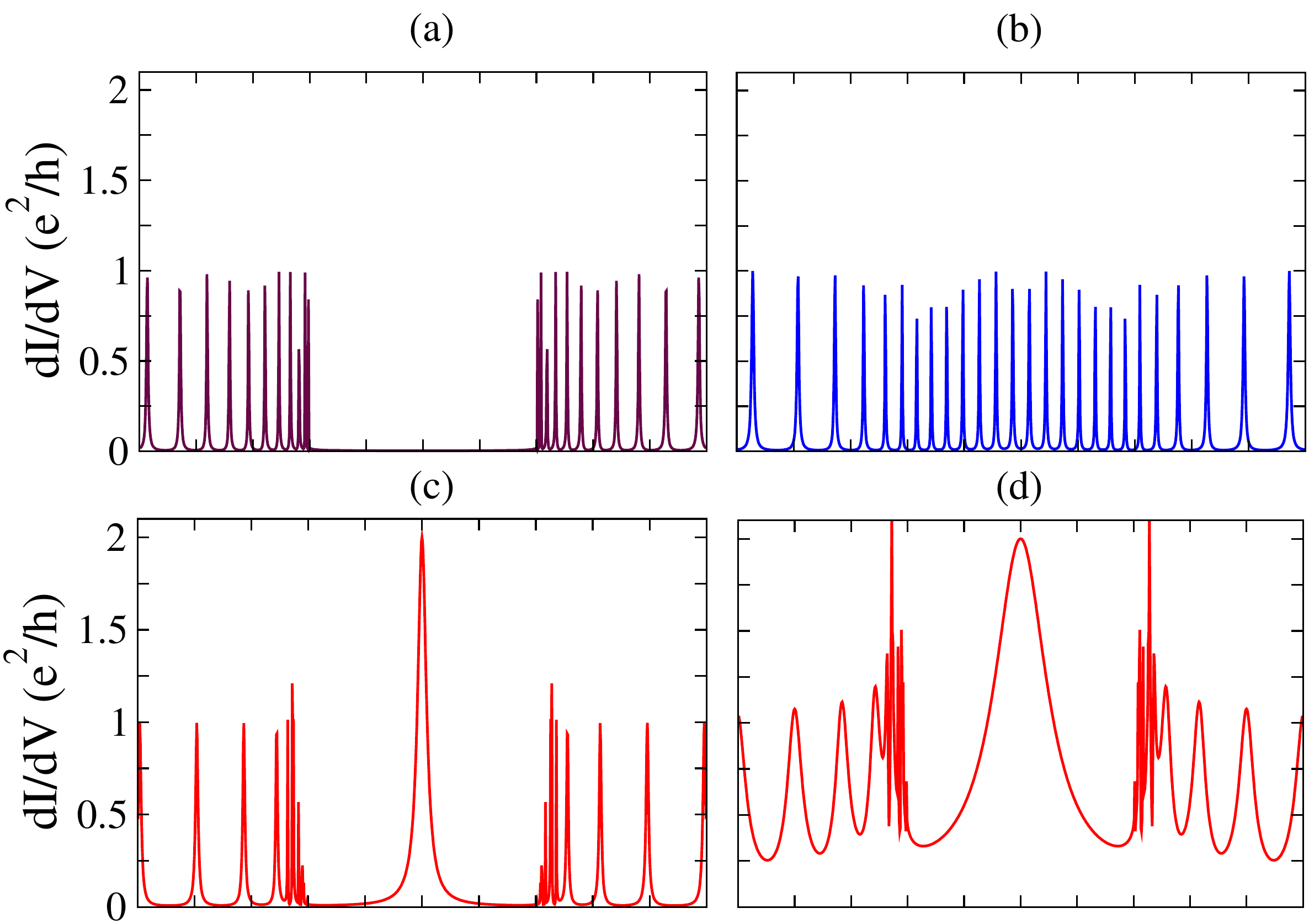,width=0.9\linewidth,clip=}\\
\epsfig{file=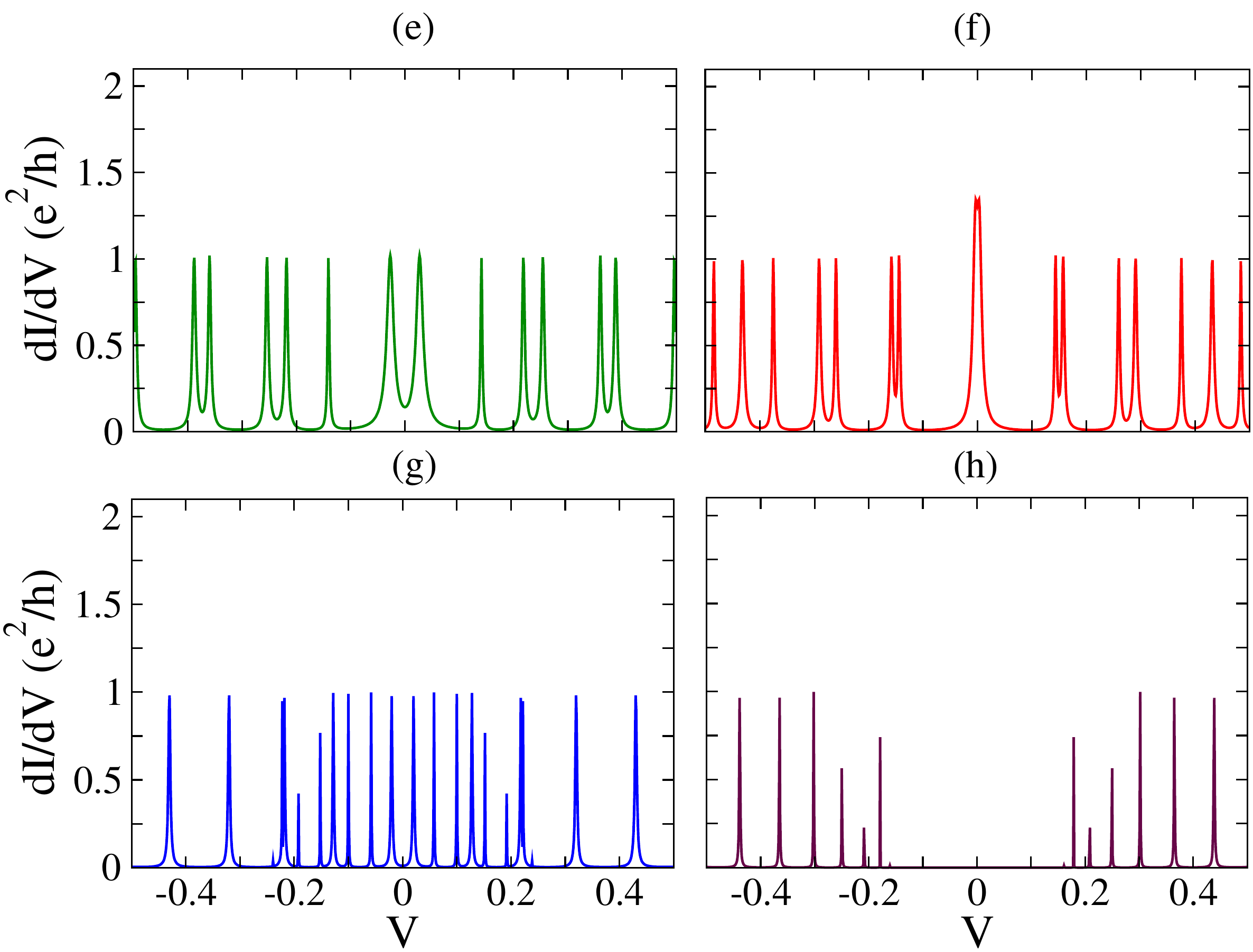,width=0.9\linewidth,clip=} 
\end{tabular}
\caption{Zero temperature $dI/dV$ \textit{vs.}~$V$ with increasing magnetic field. Everywhere $N=40$, $\gamma_{m}=\gamma=1$, $\alpha=0.2$, $\mu=0$ and $|\Delta|=0.3$. The magnetic field is chosen as follows: (a) $B=0.2$, (b) $B=0.3$, (c,d) $B=0.4$, (e) $B=1.25$, (f) $B=1.5$, (g) $B=1.9$, and (h) $B=2.1$.  Everywhere $\gamma^{\prime}_{m}=0.2$ ($m=L,R$), except $\gamma^{\prime}_{m}=0.5$ in (d). }
\label{ZBCP1} 
\end{figure}

\begin{figure}[htb]
\begin{center}
\includegraphics[width=8.0cm]{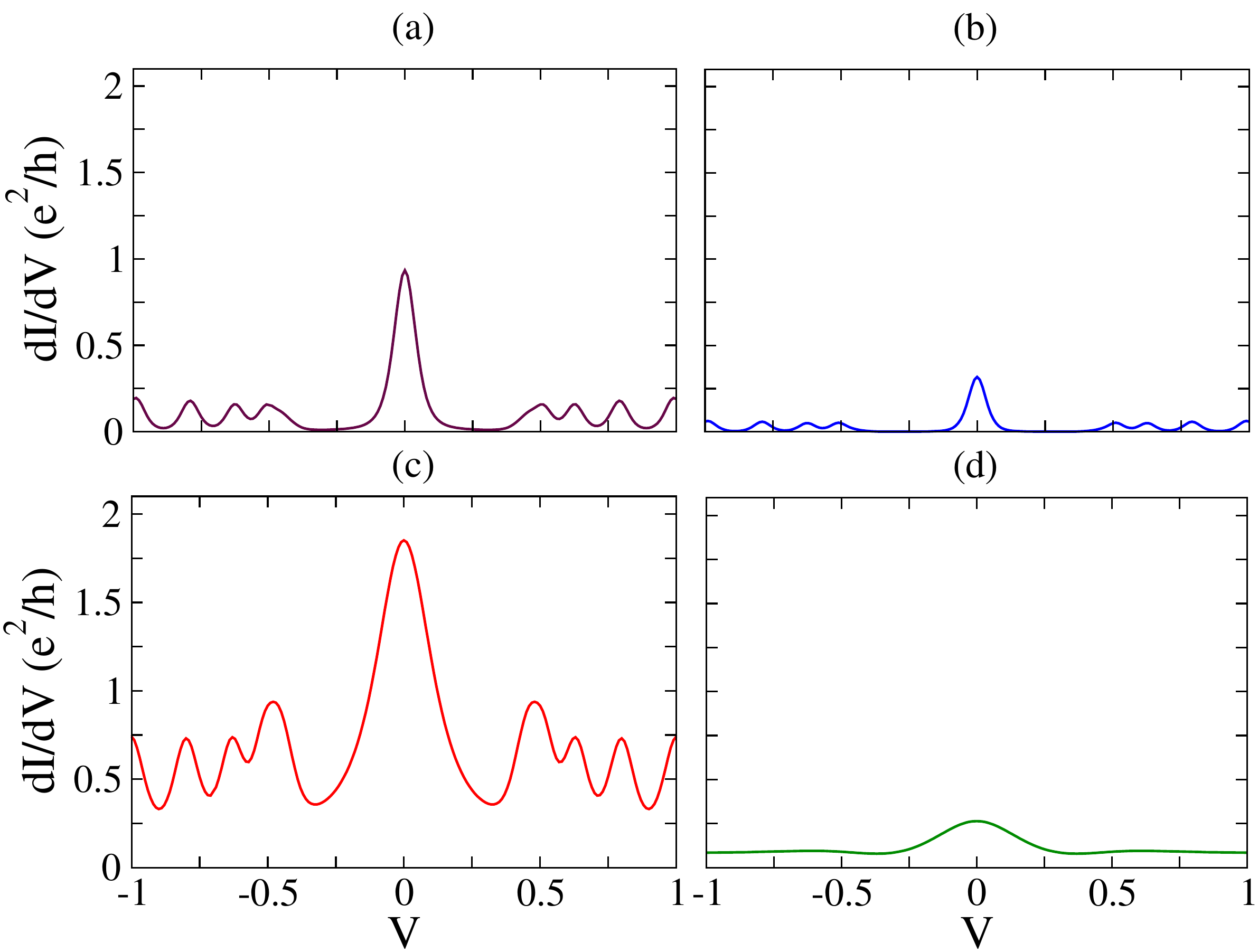}
\end{center}
\caption{Finite-temperature $dI/dV$ \textit{vs.}~$V$ at different temperatures ($k_BT$) and strengths of contacts. Everywhere $N=40$, $\gamma_{m}=\gamma=1$ (with $m=L,R$), $\mu=0$, $\alpha=0.2$, $B=0.4$ and $|\Delta|=0.3$. The rest of the parameters are as follows: (a)  $\gamma^{\prime}_{m}=0.2$, $k_BT=0.01$; (b) $\gamma^{\prime}_{m}=0.1$, $k_BT=0.01$; (c) $\gamma^{\prime}_{m}=0.5$, $k_BT=0.01$; and (d) $\gamma^{\prime}_{m}=0.2$, $k_BT=0.05$.}
\label{ZBCP0} 
\end{figure}

One can also calculate the finite temperature differential conductance from Eq.~\ref{lsCurr}, which is important to understand the recent experiments. At zero temperature and lower magnetic field (for relatively low $\mu$) the height of the ZBCP in the topologically nontrivial phase is independent of the contacts with the wire as shown in Fig.~\ref{ZBCP1} panels (c,d) for two different contact strengths. Those two plots show that the width of the ZBCP increases with increasing strength of contacts and the conductance within the pairing gap becomes also finite for stronger contacts. The latter can be understood because the dephasing induced by the baths becomes substantial with an increasing strength of the contacts and that creates finite conductance even within the superconducting pairing gap. However, at finite temperatures such that $k_BT>\Gamma$, the height of the ZBCP depends on the strength of the contacts \cite{Lin12}. When that happens, the height of ZBCP falls rapidly with increasing temperature of the baths as illustrated in Fig.~\ref{ZBCP0} panels (a,b,d). If $\Gamma>k_BT$ the effect of temperature on the height of the peak is minimal and the ZBCP remains almost like at zero temperature [see Fig.~\ref{ZBCP0}(c)].

In Fig.~\ref{ZBCP1} we show the nature of the zero-temperature differential conductance with an increasing magnetic field. When the applied field is $B<B_c$ (with $B_c=\Delta$ since $\mu=0$ in the case of the figure), there is a gap in the zero-temperature differential conductance around zero voltage for relatively weaker contacts; as is shown in Fig.~\ref{ZBCP1}(a). The gap in $dI/dV$ closes at $B=B_c$ as shown in Fig.~\ref{ZBCP1}(b). For $B$ just above $B_c$, the gap reopens and a ZBCP appears in the $dI/dV$ characteristics with its height being $2e^2/h$ at zero temperature for the ideal case; this is shown in Fig.~\ref{ZBCP1}(c). As we further increase $B$, the pairing gap separating the MBSs from the higher excitations is reduced. Here we find a splitting in the ZBCP as well as finite voltage conductance peaks inside the superconducting pairing gap. The height of the split MBS peak is $e^2/h$ and it is shown in Fig.~\ref{ZBCP1}(e). The splitting of the ZBCP disappears as we increase $B$ further, however the height of ZBCP does not retrieve to its full value of $2e^2/h$ as is shown in Fig.~\ref{ZBCP1}(f). As the field is increased even more, we find first a new closing of the gap, and finally a topologically trivial gapped superconducting phase without MBSs. These are shown in Fig.~\ref{ZBCP1}(g,h). Recently, an oscillation in the splitting of the ZBCP with increasing magnetic field has been proposed as a smoking gun for detection of the MBSs in these hybrid structures \cite{DasSarma12}. Here we find from a full microscopic transport calculation that one can expect one such oscillation in the splitting of the ZBCP with increasing field. However, we also show that the height of the ZBCP is reduced with increasing field which is an important piece of information for the experiments, and has indeed been observed recently \cite{Churchill13}. An additional important finding is the retrieval of the gap-closing phenomena at higher magnetic fields coincident with the disappearance of the MBSs, which is analogous to the lower-field scenario just before the emergence of the MBSs and reminiscent of the way the topological-nontopological quantum phase transition takes place as a function of onsite potential for a Kitaev chain \cite{Roy12}.

\begin{figure}[htb]
\begin{center}
\includegraphics[width=8.0cm]{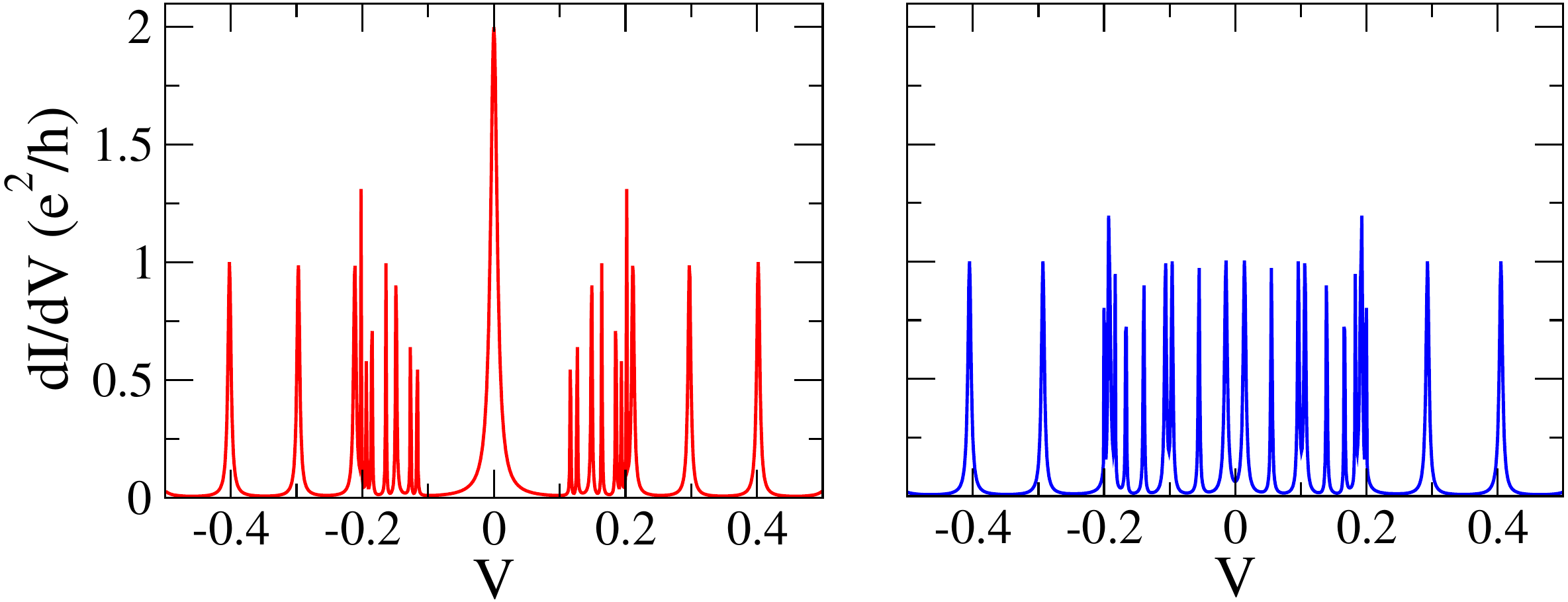}
\end{center}
\caption{Zero-temperature $dI/dV$ \textit{vs.}~$V$ with decreasing Rashba spin-orbit coupling. Everywhere $N=40$, $\gamma_{m}=\gamma=1$, $\gamma^{\prime}_{m}=0.2$ (with $m=L,R$), $\mu=0$, $B=0.4$ and $|\Delta|=0.3$. The Rashba spin-orbit coupling is chosen as follows: (a) $\alpha=0.1$, and (b) $\alpha=0.025$.}
\label{ZBCP2} 
\end{figure}

Next we study the role of the Rashba spin-orbit coupling $\alpha$ to observe the ZBCP associated with the zero-energy Majorana fermions. Usually these semiconductor-superconductor hybrid structures exhibit small Rashba splittings which is of order $\alpha k_F \sim 0.1$ meV where $k_F$ is the Fermi momentum for $\mu=0$. Therefore we look for the effect on the ZBCP of a decreasing strength of $\alpha$. The pairing gap around the ZBCP is reduced for smaller $\alpha$ as shown in Fig.~\ref{ZBCP2}(a). Further, the gap is fully closed for an even smaller value of the spin-orbit coupling and the ZBCP associated with the MBSs disappears; as shown in Fig.~\ref{ZBCP2}(b).

\begin{figure}[htb]
\begin{center}
\includegraphics[width=8.0cm]{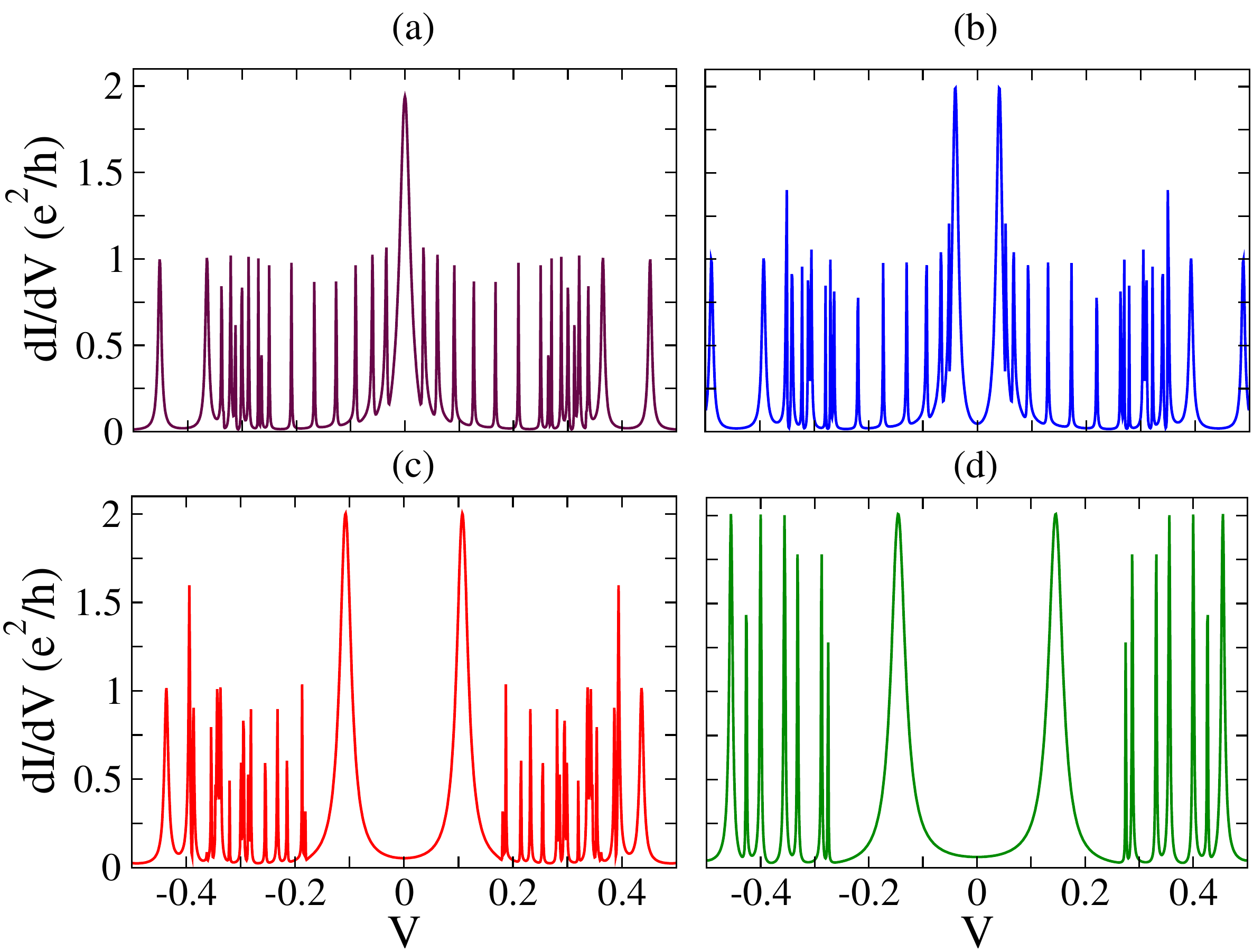}
\end{center}
\caption{Zero-temperature $dI/dV$  \textit{vs.}~$V$ with increasing $\mu$ at relatively lower values of $B$. Everywhere $N=40$, $\gamma_{m}=\gamma=1$, $\gamma^{\prime}_{m}=0.2$ (with $m=L,R$), $\alpha=0.2$, $B=0.4$ and $|\Delta|=0.3$. The gate-controlled onsite potential is as follows: (a) $\mu=0.25$, ($B_c=0.39$); (b) $\mu=0.3$, ($B_c=0.42$); (c) $\mu=0.5$, ($B_c=0.58$); and (d) $\mu=1$, ($B_c=1.04$).}
\label{ZBCP3} 
\end{figure}

The onsite potential of these hybrid structures can be tuned by applying gate voltages. However it is difficult to measure this potential in the current experiments (simultaneously, the number of channels in the nanowire is not exactly known). It has been predicted earlier that there can be ZBCP in the confined hybrid structures even when $\mu>B$ for a relatively high magnetic field \cite{Kells12}. Here we check the  zero-temperature behavior of the ZBCP at relatively lower and higher magnetic fields. At a lower fixed magnetic field $B$, there is a ZBCP when $\mu$ is such that $B>B_c \equiv \sqrt{\Delta^2+\mu^2}$ as shown in Fig.~\ref{ZBCP3}(a). Only one sub-band of the wire is occupied at this value of $\mu$ and the spectrum of the isolated wire is gapless. For a higher value of $\mu$ both the sub-bands are occupied, and then there are two MBSs coupled by $s$-wave pairing  at each end of the wire. The energy of the two MBSs at the same end of the wire in the topologically trivial phase ($B$ becomes smaller than $B_c$ for the corresponding $\mu$) is non-zero. Therefore the differential conductance shows two finite-voltage strong peaks at these values of $\mu$. All these features are shown in Fig.~\ref{ZBCP3}(b,c,d).
Interestingly, the height of the split peaks remains $2e^2/h$ which is very different from the behavior of the split peaks with increasing magnetic field at $\mu=0$ [cf.~Fig.~\ref{ZBCP1}(e)]. While the split peaks with height $2e^2/h$ are coming from the two coupled MBSs at the same end of the wire, the split peaks at $\mu=0$ with increasing $B$ are due to the overlap of two MBSs from the two ends of the nanowire in the topologically nontrivial phase. 
 
\begin{figure}
\begin{tabular}{cc}
\epsfig{file=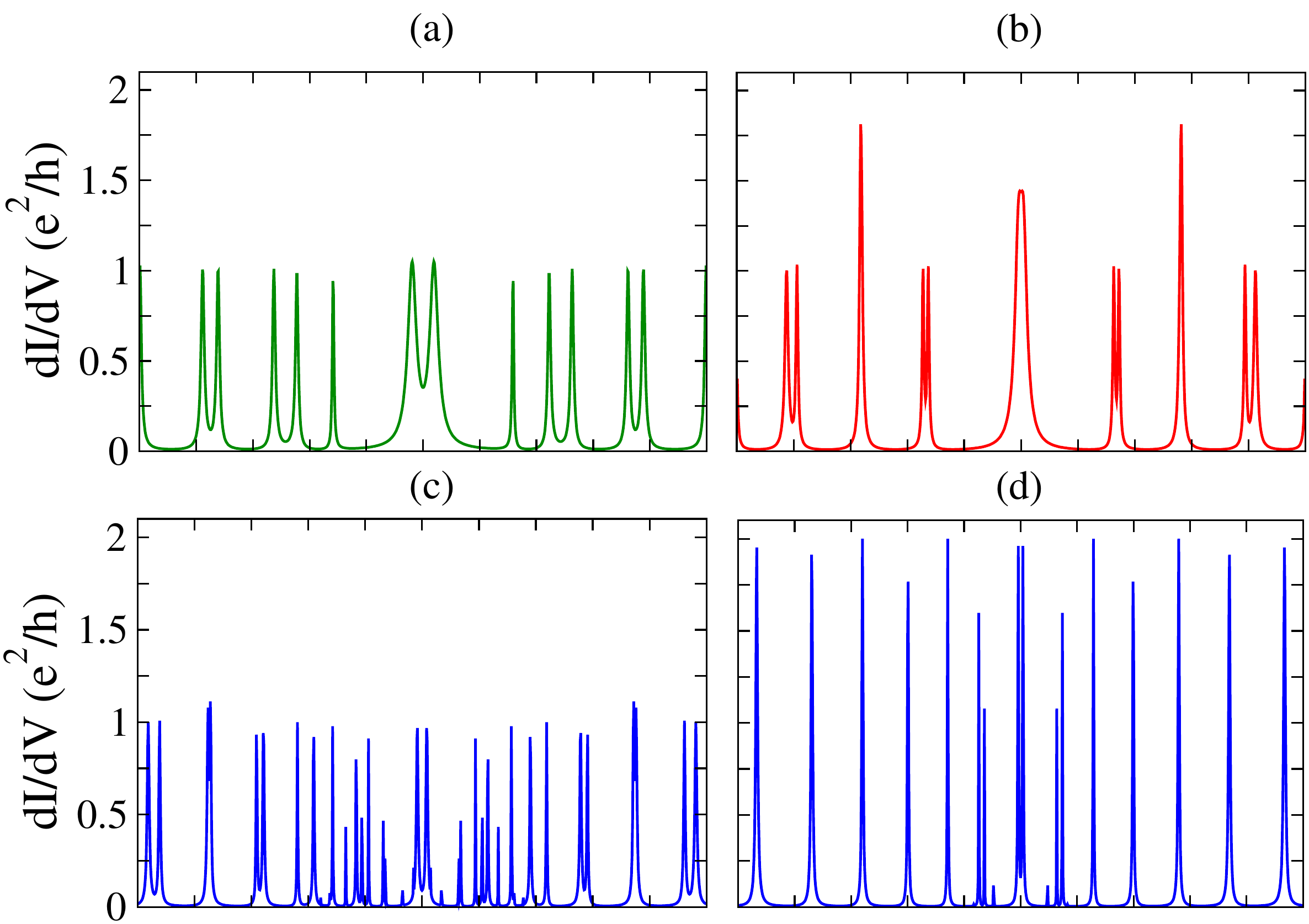,width=0.9\linewidth,clip=}\\
\epsfig{file=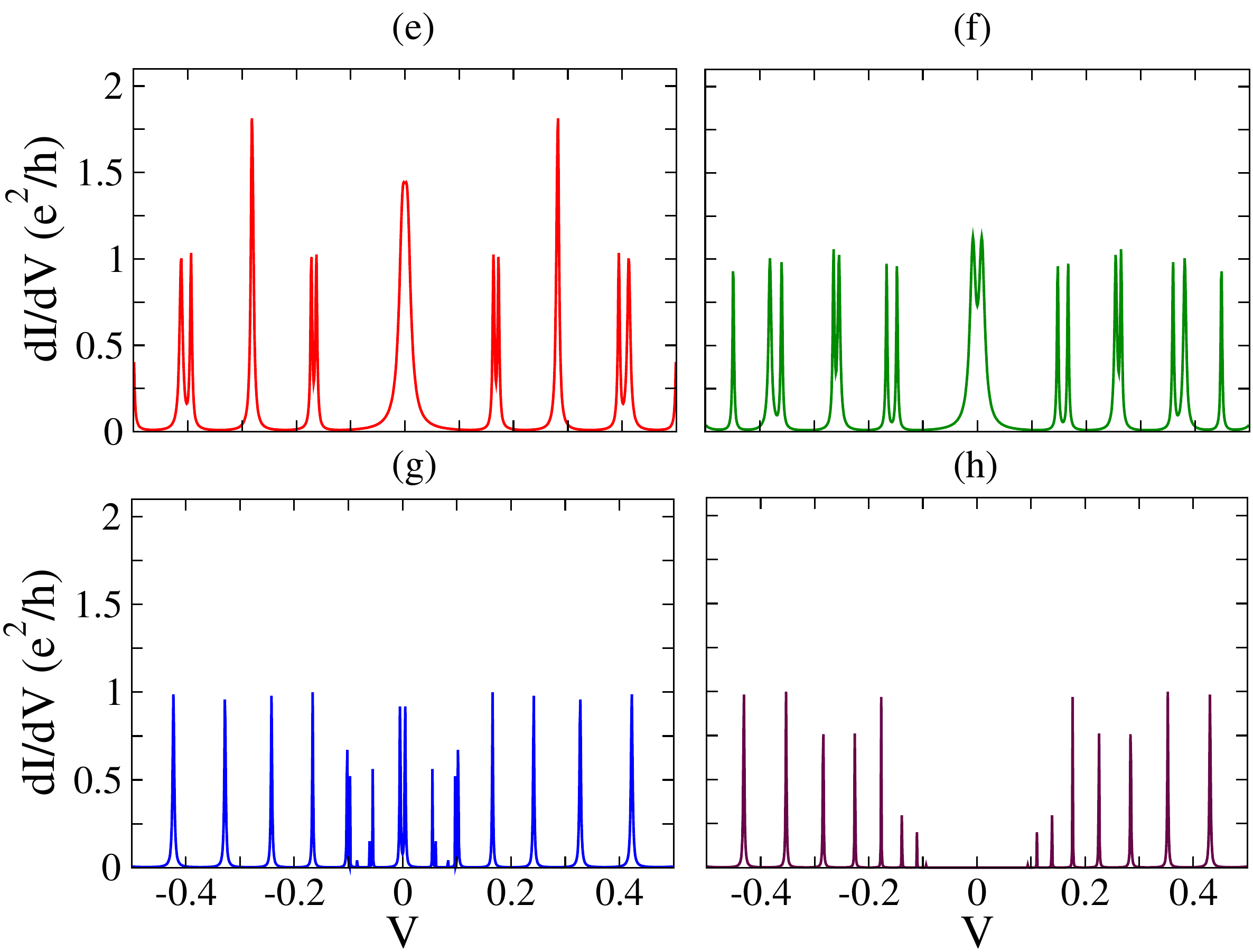,width=0.9\linewidth,clip=} 
\end{tabular}
\caption{Zero temperature $dI/dV$ \textit{vs.}~$V$ with increasing $\mu$ at larger $B$. Everywhere $N=40$, $\gamma_{m}=\gamma=1$, $\gamma^{\prime}_{m}=0.2$ ($m=L,R$), $\alpha=0.2$, $B=1$ and $|\Delta|=0.3$. The gate-controlled onsite potential is as follows: (a) $\mu=0.25$, ($B_c=0.39$); (b) $\mu=0.5$, ($B_c=0.58$); (c) $\mu=0.95$, ($B_c=1.00$); (d) $\mu=1$, ($B_c=1.04$); (e) $\mu=1.5$, ($B_c=1.53$); (f) $\mu=2.5$, ($B_c=2.52$); (g) $\mu=2.9$, ($B_c=2.92$); and (h) $\mu=3.0$, ($B_c=3.01$).}
\label{ZBCP4} 
\end{figure}

On the other hand, the behavior of the ZBCP at zero temperature with increasing $\mu$ is quite different at a relatively higher magnetic field. We start with values of $\mu$ and $B$ so that $B\gg B_c$ at that $\mu$. There the ZBCP is split and the height of the peaks is almost $e^2/h$ as shown in Fig.~\ref{ZBCP4}(a). As we increase $\mu$ the splitting of the ZBCP first disappears and then reappears. These are shown in Fig.~\ref{ZBCP4}(b,c). The values of $\Delta$, $\mu$ and $B$ in Fig.~\ref{ZBCP4}(b,c) correspond to a topologically nontrivial phase. When $\mu$ reaches a value where $B$ becomes exactly equal to $B_c$ of the corresponding $\mu$, the superconducting pairing gap closes [see Fig.~\ref{ZBCP4}(d)]. For an even larger $\mu$ the hybrid structure would be in principle in the topologically trivial phase for that $B$ as now $B<B_c$ at that $\mu$. However, we find that the ZBCP reappears again with an increasing $\mu$ but the height is smaller than $2e^2/h$, which is shown in Fig.~\ref{ZBCP4}(e). As we further increase the onsite potential, the ZBCP splits and finally disappears with another gap closing step.  For a very large value of $\mu$ the spectrum is fully gapped at the given field. It is interesting to compare the emergence of ZBCP in the topologically trivial phase with the findings of Ref.\onlinecite{Kells12} for single and multichannel Majorana wires. 

We have also found that the height of the ZBCP at high magnetic fields, both in the topologically nontrivial and trivial phases, is susceptible to the strength of tunnel contacts even at zero temperature. The height increases for a stronger tunnel contact with the baths (lower or smoother barrier potential) even at zero temperature. This behavior is different from the nature of ZBCP in the topologically nontrivial phase at a relatively smaller magnetic field [as shown in Fig.~\ref{ZBCP1} panels (c,d)]. The splitting oscillation of the ZBCP with changing gate-controlled onsite energy has been observed in Ref.~\onlinecite{Churchill13} where the height of the ZBCP is much smaller than $2e^2/h$ even at the lowest temperatures accessible to experiments. Therefore, we suspect that the recent experiments have probed this regime at high $B$ and $\mu$. However, the $p$-wave component of the induced superconducting state is very small for these parameters and the Majorana end states are not well protected from the bulk excitations due to a reduced $p$-wave component of the induced pairing gap.

\section{Discussion and Prospects}
In summary, we have provided a complete theory of both linear and nonlinear transport in realistic microscopic models of hybrid semiconductor-superconductor heterostructures with spin-orbit coupling and have shown the presence of topological superconducting states as revealed by tunneling transport.
This is manifest in the current-voltage characteristics via a ZBCP with a distinctive variation as a function of temperature and various experimentally tunable parameters. 
We used and provided a detail derivation of the LEGF method, which we further extended from our previous work \cite{Roy12} to include the elements of spin-orbit coupling and applied magnetic fields. 
A nonmonotonic dependence of the height of the ZBCP with the applied magnetic field and the onsite potential constitutes a characteristic feature of these systems that cannot be explained away by the presence of spurious conduction channels (due to the presence of disorder or to other features of the nanofabrication of the heterostructures). 
Moreover, the predicted alternating splitting and width modulation in the ZBCP with an increasing magnetic field beyond $B_c$ is a feature that cannot be generically obtained via a two-level system mechanism (cf.~Ref.~\onlinecite{Lee12}), not even with a Zeeman-field tuning of single-particle localized levels into degeneracy (similar to the scenario for the so called singlet-triplet Kondo effect in quantum dots).
The above, combined with the characteristic feature of the coordinated opening and closing of the superconducting gap as the MBS's ZBCP first appears (and then again when it disappears), constitutes a reasonably individualized scenario that would present a compelling case for the observation of Majorana states.
We have further shown how the increase of the tunnel coupling acts to modify the height of the ZBCP at finite temperatures and controls the appearance of sub-gap conductance.
Our modeling of the contacts is more realistic than in more simplified models prevalent in the literature and as a result we can meaningfully compare features arising due the the presence of MBSs versus other quasiparticle excitations.
How these characteristic features are reflected in other types of transport experiments involving interferometry in Josephson-junction geometries is an important question for further theoretical and experimental study. 

\acknowledgements
One of us (DR) is indebted to S.~Tewari and J.~D.~Sau for valuable discussions.
The authors acknowledge financial support from the University of Cincinnati and DR also acknowledges support from the U.S. Department of Energy through the LANL/LDRD Program for this work. CJB and NS acknowledge the Kavli Institute for Theoretical Physics (NSF supported under Grant No. PHY05-51164) for its hospitality while this work was in progress.

\appendix
\begin{widetext}
\section{Generalized Quantum Langevin Equations}
\label{GQLE}
The generalized quantum Langevin equations for the wire operators in the Majorana basis are explicitly given by the following two expressions:
{\allowdisplaybreaks\begin{eqnarray}
\dot{c}_{A,l,\sigma}&=&\f{2(\mu-\gamma)}{\hbar}c_{B,l,\sigma}+ \f{2(B+\Delta)}{\hbar}\delta_{\sigma,\uparrow}c_{B,l,\downarrow}+\f{2(B-\Delta)}{\hbar}\delta_{\sigma,\downarrow}c_{B,l,\uparrow}-\f{\gamma}{\hbar}c_{B,l+1,\sigma}-\f{\gamma}{\hbar}c_{B,l-1,\sigma}+\f{\alpha}{\hbar}\delta_{\sigma,\downarrow}c_{B,l+1,\uparrow}\nn\\&+&\f{\alpha}{\hbar}\delta_{\sigma,\uparrow}c_{B,l-1,\downarrow}-\f{\alpha}{\hbar}\delta_{\sigma,\downarrow}c_{B,l-1,\uparrow}-\f{\alpha}{\hbar}\delta_{\sigma,\uparrow}c_{B,l+1,\downarrow}+\delta_{l,1}\big(-i\n_{L,\sigma}-i\int_{t_0}^\infty dt' \Sigma^+_L (t-t')\f{1}{2}(c_{A,1,\sigma}(t')+ic_{B,1,\sigma}(t'))\nn\\&+&i\n^{\dag}_{L,\sigma}+ i \int_{t_0}^\infty dt' \Sigma^-_L (t-t') \f{1}{2}(c_{A,1,\sigma}(t')-ic_{B,1,\sigma}(t'))\big)+\delta_{l,N}\big(-i\n_{R,\sigma} -i \int_{t_0}^\infty dt' \Sigma^+_R (t-t') \nn\\&& \times \f{1}{2}(c_{A,N,\sigma}(t')+ic_{B,N,\sigma}(t'))+i\n^{\dag}_{R,\sigma}+ i \int_{t_0}^\infty dt' \Sigma^-_R (t-t') \f{1}{2}(c_{A,N,\sigma}(t')-ic_{B,N,\sigma}(t'))\big),\label{gle1}\\
\dot{c}_{B,l,\sigma}&=&-\f{2(\mu-\gamma)}{\hbar}c_{A,l,\sigma}- \f{2(B-\Delta)}{\hbar}\delta_{\sigma,\uparrow}c_{A,l,\downarrow}-\f{2(B+\Delta)}{\hbar}\delta_{\sigma,\downarrow}c_{A,l,\uparrow}+\f{\gamma}{\hbar}c_{A,l+1,\sigma}+\f{\gamma}{\hbar}c_{A,l-1,\sigma}-\f{\alpha}{\hbar}\delta_{\sigma,\downarrow}c_{A,l+1,\uparrow}\nn\\&-&\f{\alpha}{\hbar}\delta_{\sigma,\uparrow}c_{A,l-1,\downarrow}+\f{\alpha}{\hbar}\delta_{\sigma,\downarrow}c_{A,l-1,\uparrow}+\f{\alpha}{\hbar}\delta_{\sigma,\uparrow}c_{A,l+1,\downarrow}+\delta_{l,1}\big(-\n_{L,\sigma} -\int_{t_0}^\infty dt' \Sigma^+_L (t-t')\f{1}{2}(c_{A,1,\sigma}(t')+ic_{B,1,\sigma}(t'))\nn\\&-&\n^{\dag}_{L,\sigma}- \int_{t_0}^\infty dt' \Sigma^-_L (t-t') \f{1}{2}(c_{A,1,\sigma}(t')-ic_{B,1,\sigma}(t'))\big)+\delta_{l,N}\big(-\n_{R,\sigma} -\int_{t_0}^\infty dt' \Sigma^+_R (t-t')\nn\\&& \times \f{1}{2}(c_{A,N,\sigma}(t')+ic_{B,N,\sigma}(t'))-\n^{\dag}_{R,\sigma}- \int_{t_0}^\infty dt' \Sigma^-_R(t-t') \f{1}{2}(c_{A,N,\sigma}(t')-ic_{B,N,\sigma}(t'))\big).\label{gle2}
\end{eqnarray}}

\section{Steady-state solution of the operators on the nanowire}
\label{GreenFn}
Expressions for the frequency functions entering the definition of the Green's function of the full system consisting of the nanowire and the baths, $G^+(\om)$ (cf.~\cite{DharSen06}) and used in the steady state solution of the Majorana operators:
{\allowdisplaybreaks\bea
\Phi_{lm}(\om)&=&\omega~\delta_{lm}-\f{2i(\mu-\gamma)}{\hbar}\delta_{l,m-2}\delta_{\rm mod(m+1,4),0}-\f{2i(\Delta+B)}{\hbar}\delta_{l,m-3}\delta_{\rm mod(m,4),0}+\f{i\gamma}{\hbar}\delta_{l,m-6}\delta_{\rm mod(m+1,4),0}\nn\\ &+&\f{i\alpha}{\hbar}\delta_{l,m-7}\delta_{\rm mod(m,4),0}-\f{2i(\mu-\gamma)}{\hbar}\delta_{l,m-2}\delta_{\rm mod(m,4),0}+\f{2i(\Delta-B)}{\hbar}\delta_{l,m-1}\delta_{\rm mod(m+1,4),0}+\f{i\gamma}{\hbar}\delta_{l,m-6}\delta_{\rm mod(m,4),0}\nn\\ &-&\f{i\alpha}{\hbar}\delta_{l,m-5}\delta_{\rm mod(m+1,4),0}+\f{2i(\mu-\gamma)}{\hbar}\delta_{l,m+2}\delta_{\rm mod(l+1,4),0}-\f{2i(\Delta-B)}{\hbar}\delta_{l,m+1}\delta_{\rm mod(l+1,4),0}-\f{i\gamma}{\hbar}\delta_{l,m-2}\delta_{\rm mod(m-1,4),0}\nn\\ &-&\f{i\alpha}{\hbar}\delta_{l,m-3}\delta_{\rm mod(m-2,4),0}+\f{2i(\mu-\gamma)}{\hbar}\delta_{l,m+2}\delta_{\rm mod(l,4),0}+\f{2i(\Delta+B)}{\hbar}\delta_{l,m+3}\delta_{\rm mod(l,4),0}-\f{i\gamma}{\hbar}\delta_{l,m-2}\delta_{\rm mod(l,4),0}\nn\\ &+&\f{i\alpha}{\hbar}\delta_{l,m-1}\delta_{\rm mod(l,4),0}+\f{i\gamma}{\hbar}\delta_{l,m+2}\delta_{\rm mod(l-1,4),0}-\f{i\alpha}{\hbar}\delta_{l,m+1}\delta_{\rm mod(m,4),0}+\f{i\gamma}{\hbar}\delta_{l,m+2}\delta_{\rm mod(m,4),0}+\f{i\alpha}{\hbar}\delta_{l,m+3}\delta_{\rm mod(m+1,4),0}\nn\\&-&\f{i\gamma}{\hbar}\delta_{l,m+6}\delta_{\rm mod(l+1,4),0}+\f{i\alpha}{\hbar}\delta_{l,m+5}\delta_{\rm mod(l+1,4),0}-\f{i\gamma}{\hbar}\delta_{l,m+6}\delta_{\rm mod(l,4),0}-\f{i\alpha}{\hbar}\delta_{l,m+7}\delta_{\rm mod(l,4),0}, \\
A_{lm}(\om)&=&- \Se^+_L(\om)\delta_{l,m}(\delta_{l,1}+\delta_{l,2}+\delta_{l,3}+\delta_{l,4})- \Se^+_R(\om)\delta_{l,m}(\delta_{l,4N-3}+\delta_{l,4N-2}+\delta_{l,4N-1}+\delta_{l,4N}),\\
\tilde{h}_m(\om)&=&\big[\tn_{L,\uparrow}(\om)-\tn^{\dag}_{L,\uparrow}(-\om)]\delta_{m,1}+\big[\tn_{L,\downarrow}(\om)-\tn^{\dag}_{L,\downarrow}(-\om)]\delta_{m,2}-i\big[\tn_{L,\uparrow}(\om)+\tn^{\dag}_{L,\uparrow}(-\om)]\delta_{m,3}-i\big[\tn_{L,\downarrow}(\om)+\tn^{\dag}_{L,\downarrow}(-\om)]\delta_{m,4}\nn\\&+&\big[\tn_{R,\uparrow}(\om)-\tn^{\dag}_{R,\uparrow}(-\om)]\delta_{m,4N-3}+\big[\tn_{R,\downarrow}(\om)-\tn^{\dag}_{R,\downarrow}(-\om)]\delta_{m,4N-2}-i\big[\tn_{R,\uparrow}(\om)+\tn^{\dag}_{R,\uparrow}(-\om)]\delta_{m,4N-1}\nn\\&-&i\big[\tn_{R,\downarrow}(\om)+\tn^{\dag}_{R,\downarrow}(-\om)]\delta_{m,4N}.\label{Noise}
\eea}
\end{widetext}

\end{document}